\documentclass[conference]{IEEEtran}

\usepackage{amsmath, amssymb, amsfonts, bm}

\usepackage{graphicx}
\usepackage{float}      
\usepackage{comment}
\usepackage[vlined, ruled, linesnumbered]{algorithm2e}
\SetKwInput{KwInput}{Input}
\SetKwInput{KwOutput}{Output}
\SetKwRepeat{Do}{do}{until}

\makeatother

\usepackage{xcolor}
\usepackage{listings}
\usepackage{balance}

\lstdefinelanguage{Matlab}{
  morekeywords={break,case,catch,continue,else,elseif,end,for,function,
    global,if,otherwise,persistent,return,switch,try,while},
  sensitive=true,
  morecomment=[l]\%,
  morestring=[m]',
}
\usepackage{amsthm}
\theoremstyle{plain}
\newtheorem{theorem}{Theorem}[section]

\newtheorem{proposition}[theorem]{Proposition}
\newtheorem{corollary}[theorem]{Corollary}
\theoremstyle{definition}
\newtheorem{definition}[theorem]{Definition}
\newtheorem{example}{Example}
\newtheorem{remark}{Remark}

\makeatletter
\def\th@plain{%
  \thm@notefont{}%
  \itshape
  \normalfont   
}
\makeatother

\usepackage{cite}

\usepackage{tikz}
\usetikzlibrary{positioning,shapes,arrows.meta,calc}
\usetikzlibrary{decorations.pathreplacing}

\tikzset{
  >={Latex[length=3mm]},
  block/.style={draw, thick, rectangle, minimum width=2.8cm, minimum height=1.2cm, align=center},
  filter/.style={draw, thick, trapezium, trapezium left angle=70, trapezium right angle=110,
                 minimum width=3.0cm, minimum height=1.2cm, align=center},
  mixer/.style={circle, draw, thick, minimum size=8mm, inner sep=0pt,
                path picture={
                  \draw[]
                    (-2.4mm,-2.4mm)--(2.4mm,2.4mm)
                    (-2.4mm,2.4mm)--(2.4mm,-2.4mm);
                }},
  conn/.style={thick, -{Latex}}
}

\DeclareFontFamily{U}{mathx}{\hyphenchar\font45}
\DeclareFontShape{U}{mathx}{m}{n}{
      <5> <6> <7> <8> <9> <10>
      <10.95> <12> <14.4> <17.28> <20.74> <24.88>
      mathx10
      }{}
\DeclareSymbolFont{mathx}{U}{mathx}{m}{n}
\DeclareFontSubstitution{U}{mathx}{m}{n}
\DeclareMathAccent{\widecheck}{0}{mathx}{"71}

\DeclareFontFamily{U}{mathx}{\hyphenchar\font45}
\DeclareFontShape{U}{mathx}{m}{n}{<-> mathx10}{}
\DeclareSymbolFont{mathx}{U}{mathx}{m}{n}
\DeclareMathAccent{\widebar}{0}{mathx}{"73}

\newfont{\Bd}{msbm10 at 12 truept}
\newfont{\Sc}{eusm10 at 12 truept}

\newcommand\fH[1]{\sbox0{#1}\dimen0=\ht0 \advance\dimen0 -1ex
  \sbox2{\'{}}\sbox2{\raise\dimen0\box2}%
  {\ooalign{\hidewidth\kern.1em\copy2\kern-.5\wd2\box2\hidewidth\cr\box0\crcr}}}

\def\0{{\bf 0}}
\def\1{{\bf 1}}

\def\c{{\bf c}}

\def\m{{\bf m}}

\def\u{{\bf u}}

\def\y{{\bf y}}

\def\A{{\bf A}}
\def\B{{\bf B}}

\def\E{{\bf E}}
\def\F{{\bf F}}
\def\G{{\bf G}}
\def\H{{\bf H}}

\def\M{{\bf M}}

\def\0{{\bf 0}}
\def\P{{\bf P}}

\def\W{{\bf W}}

\def\mbF{{ \mathbb{F} }}

\def\mcC{{ \mathcal{C} }}

\def\mcI{{ \mathcal{I} }}

\def\mcN{{ \mathcal{N} }}

\def\mcR{{ \mathcal{R} }}

\def\pod{{\mathsf{PED}}}
\def\scl{{\mathsf{SCL}}}
\def\sc{{\mathsf{SC}}}

\def\polar{p}
\def\BLBC{b}
\def\rowspan{\text{rowspan}}
\def\Sym{\text{Sym}}
\def\Aut{\text{Aut}}

\makeatletter
\def\ifundefined{\@ifundefined}
\makeatother

\title{
    Parallel Decoding of Binary Linear Block Codes via Equivalent Polar Transformation Class
  }

\author{Pin-Jing Li and Yu-Chih Huang \\
Institute of Communications Engineering, National Yang Ming Chiao Tung University\\
\{ouo.ee11@nycu.edu.tw, jerryhuang@nycu.edu.tw\}
}

\begin{document}
\maketitle
\begin{abstract}
    A parameterized universal decoding framework based on polar 
    transformations was recently proposed to enable polar-style 
    decoding of general binary linear block codes (BLBCs).
    However, existing parallelization methods for polar codes 
    cannot be directly applied to this framework. 
    The key obstacle is that these methods rely on static frozen-set structures, 
    which are incompatible with the dynamic frozen constraints induced by polar transformations.

    {To address this challenge, 
    we revisit code automorphisms from a new perspective: 
    instead of applying them to the codeword space, 
    we let them act on the transformation parameter itself.  }
    We show that the BLBC automorphism group 
    induces equivalence polar transformation classes. 
    Crucially, all transformations within an equivalence class 
    share the same polar subcode structure, eliminating the need for separate decoder designs.

    This insight enables a novel parallel decoding framework 
    termed polar ensemble decoding (PED). 
    By decoding multiple equivalent transformations simultaneously, 
    PED exploits transformation diversity while maintaining decoder compatibility. 
    Simulation results on extended BCH and extended Golay codes demonstrate that PED 
    achieves near maximum-likelihood performance while significantly 
    reducing the decoding latency compared to successive 
    cancellation list (SCL) decoding.

\end{abstract}

\section{Introduction}
Binary linear block codes (BLBCs) constitute the foundation of contemporary digital communication systems \cite{lin2001ECC}. 
Different code families 
offer distinct trade-offs in decoding complexity, 
minimum-distance guarantees, and asymptotic performance. 
As no single code simultaneously optimizes all of these criteria, 
modern standards increasingly adopt composite coding architectures for different uses.
A representative example is the 5G New Radio (NR) standard, 
adopting LDPC codes for data channels and polar codes for control channels. 
While this design achieves near-capacity performance across diverse scenarios, 
it requires multiple dedicated decoders at the receivers, 
leading to increased hardware complexity and implementation cost.
A universal decoder capable of handling multiple code families 
within a single architecture could significantly simplify the design. 

This has motivated the design of the parameterized universal BLBC decoding framework~\cite{lin2020unidec}. 
The authors embed arbitrary BLBCs in polar subcodes with dynamic frozen constraints~\cite{trifonov2013df}, 
enabling polar decoding for general BLBCs.
Subsequently~\cite{lin2025unidec}, 
the authors extended the framework to 
BLBCs of arbitrary code lengths.
This line of work has also been extended to U-UV codes~\cite{chenli2025U_UV} 
and non-binary codes~\cite{chenli2025nonbin}. 
However, as these extensions rely on SCL decoding, 
they remain inheritantly sequential and the 
parallelizations remain unexplored.

{
Several automorphism ensemble decoding (AED) schemes have been developed 
for specific code families to enable parallel soft-decision decoding,
including methods for Reed-Muller codes~\cite{geiselhart2021RM_AED},
polar codes~\cite{mondelli2014polar_RM,geiselhart2021Aut_polar,kamenev2019perm_polar},
and extended BCH (eBCH) codes~\cite{han2025eBCH_RM}.
}
In contrast, classical uses of automorphisms in algebraic coding theory, 
such as MacWilliams' permutation decoding~\cite{macwilliams1964perm}, 
were confined to hard-decision error shifting and did not provide soft-decision diversity.

Thus, existing parallel soft-decision decoding schemes remain 
{\it code-family-dependent} and rely on problem-specific constructions.
To address this, 
we proposed \textit{polar ensemble decoding} ($\pod$) 
based on the polar transformation framework~\cite{lin2020unidec}.

By applying the automorphism group to the \textit{transformation parameter}
rather than the codeword space,
we obtain equivalent polar transformations
with a \textit{shared polar subcode structure}
yet provide decoding diversity under SC-style decoding.
This enables parallelism with flexible tradeoffs among latency, complexity, and performance. 
We further analyze the LTA-type transformations within an equivalence class and derive a probabilistic bound showing that uniformly sampled elements are unlikely to exhibit LTA-type SC-invariant redundancy.
Simulation results on eBCH and extended Golay codes demonstrate the 
near maximum-likelihood performance 
of our proposed $\pod$ framework,
while significantly reducing decoding latency compared with SCL decoding.

The remainder of this paper is organized as follows. 
Section~\ref{sec:background} reviews polar codes, polar subcodes, and polar transformations of BLBCs~\cite{lin2020unidec}. 
Section~\ref{sec:POD} characterizes the algebraic properties of polar transformation,
which allow us to identify the equivalent classes of the polar transformation and 
analyze the LTA-type SC-invariant transforms.
Simulation results and conclusions are presented in Sections~\ref{sec:simulation} and~\ref{sec:conclude}, respectively.

\section{Background}\label{sec:background}

Throughout the paper, we denote 
$\mcN(\cdot)$ and $\mcR(\cdot)$ as the nullspace and rowspace of a matrix.
The operator $\text{RREF}(\cdot)$ reduces a matrix to its unique row-reduced echelon form.
We denote the symmetric group of $n$ elements as $\text{Sym}(n)$.

\subsection{Conventional Polar Codes~\cite{arikan2009channel}}

Consider a polar code of blocklength $n=2^m$ with the generator matrix 
$\G_\polar = \B_m\F^{\otimes m}$, 
where $\F=\bigl[\begin{smallmatrix}1&0\\1&1\end{smallmatrix}\bigr]$ is the Ar\i kan kernel 
and $\B_m$ is a bit-reversal permutation matrix. 
The kernel recursively transforms identical physical channels into synthetic subchannels with polarized reliabilities.
To transmit a $k$-bit message $\mathbf{m} \in \mathbb{F}_2^k$, the encoding vector $\mathbf{u} \in \mathbb{F}_2^n$ 
is constructed by placing the message on the $k$ most reliable channels $\mcI$, $\mathbf{u}_{\mathcal{I}} = \mathbf{m}$.
The remaining entries are frozen to zero, i.e., $\mathbf{u}_{\mathcal{I}^c} = \mathbf{0}$. 
The codeword is generated as $\mathbf{c} = \mathbf{u}\mathbf{G}_p.$

The 
successive cancellation (SC) decoding algorithm~\cite{arikan2009channel} for polar codes  
sequentially estimates each information bit $u_i$ 
given the received vector $\y$ and prior estimates $\hat{u}_{1}^{i-1}$.
To combat the error propagation inherent in SC, 
SCL decoding~\cite{tal2015list} tracks the $L$ most probable paths,
approaching ML performance as the list size increases under finite blocklength.

\subsection{Polar Subcode~\cite{trifonov2013df}}

Since the matrix $\G_\polar$ is invertible, 
any vector $\c$ of length $n = 2^m$ can be represented as
\(
\c = \u \G_\polar.
\)
The vector $\c$ belongs to a binary linear block code $\mcC_{\BLBC}$ 
with generator matrix $\G_{\BLBC}\in\mbF_2^{k\times n}$ and 
parity-check matrix $\H_{\BLBC}\in\mbF_2^{n-k\times n}$ when
\begin{equation}
\c \in \mcN(\H_\BLBC^T) 
\Leftrightarrow \u \in \mcN(\G_\polar\H_\BLBC^T)
\Leftrightarrow \u \in \mcR(\G_\BLBC\G_\polar^{-1}).
\end{equation}
This can be viewed as a polar subcode with the encoding vector $\u$ subjects to the dynamic frozen 
constraint $\W := \G_\BLBC\G_\polar^{-1}$. 
This indeed makes arbitrary BLBC compatible with a polar-style decoder.
However, the information set is fixed once the BLBC is determined, 
which may place message bits on unreliable channels and thus lead to a poor reliability profile.

\subsection{Universal Polar Decoding for BLBCs~\cite{lin2020unidec,lin2025unidec}}
To address the frozen set limitation,
Lin \textit{et al.}~\cite{lin2020unidec} introduced the 
dynamic frozen-bit design flexibility by manipulating invertible matrices $\G_p$ and permutations. 

\begin{proposition}(\textit{Polar Transformation~\cite{lin2020unidec}.})
\label{prop:PD}
For an $(n,k)$ BLBC $\mathcal{C}_b$ and a permutation matrix $\mathbf{P}$, 
there exists a polar subcode $\mathcal{C}_p$ with dynamic frozen-bits 
such that the mapping between $\mathbf{c}_b\in \mathcal{C}_b$ 
and $\mathbf{c}_p\in\mathcal{C}_p$ is given by 
$\mathbf{c}_b = \mathbf{c}_p \mathbf{P}^{-1}$.
\end{proposition}

\begin{IEEEproof}
 Consider $\W_\P := \G_b\P\G_p^{-1}$. For any $\c_b \in \mcC_b$, there exists $\m\in \mbF_2^k$ such that 
\begin{align}
  \c_b &= \m\G_b = \m\G_b(\P\G_p^{-1}\G_p\P^{-1}) =\m\W_\P\G_p\P^{-1}.
\end{align}
Now consider $\mathbf{E}\in\text{GL}_k(\mbF_2)$ the elimination matrix reducing 
$\W_\P$ to RREF. Let $\M_\P = \E\W_\P$, then 
\begin{align}
  \c_b & = \m(\mathbf{E}^{-1}\mathbf{E}) \W_\P\G_p\P^{-1} = (\m\mathbf{E}^{-1})(\mathbf{E} \W_\P)\G_p\P^{-1} \nonumber\\
  & = (\mathbf{m}_p \mathbf{M}_{\P} \mathbf{G}_p)\P^{-1} = \c_p\P^{-1}.
\end{align}
If we consider $\M_\P$ as the dynamic frozen matrix, then 
$\mathbf{c}_p \triangleq \mathbf{m}_p \mathbf{M}_{\P} \mathbf{G}_p$
encodes
the transformed message $\mathbf{m}_p \triangleq \mathbf{m}\mathbf{E}^{-1}$ into a polar subcode.
The row-reduced form $\M_\P$ 
 preserves the row space of $\W_\P$ 
 while yielding a dynamic frozen constraint compatible with SC-style decoding.
\end{IEEEproof}

This polar transformation allows the use of 
polar decoders on general BLBCs 
with frozen-bit design flexibility. 

{
Even though polar codes possess the lower triangular affine (LTA) automorphism group~\cite{bardet2016algebraic_polar},
this does not benefit us for a direct parallelization, since  
it is invariant under SC-style decoding.
Researchers have explored specific polar subcodes 
with other automorphisms~\cite{geiselhart2021Aut_polar,kamenev2019perm_polar}, 
but these typically require information set designs 
incompatible with the dynamic frozen constraints $\M_\P$ from polar transformation. 


Although BLBCs with non-power-of-two block lengths may still exploit the polar transformation structure through multi-kernel polar codes, such constructions are restricted to lengths factorizable by the available kernel sizes and require tailored kernel designs. In the subsequent work~\cite{lin2025unidec}, pruning and shortening operations further extended the framework to arbitrary code lengths. Since these additional design freedoms substantially enlarge the search space, the Bhattacharyya-sum optimization was addressed via stochastic search. Although the resulting framework improves decoding performance over existing universal soft-decoding methods, its algebraic characterization and parallelization remain unexplored.

}

\section{Parallelized Polar Decoding for BLBCs}\label{sec:POD} 

Building on the transformation framework in \cite{lin2020unidec,lin2025unidec}, 
one can naturally envision a parallel decoding strategy that employs multiple 
transformations and decodes them simultaneously. 
However, such parallelization has not been explored, 
primarily because a large set of suitable transformation parameters is difficult to obtain.
In this section, we algebraically characterize the polar transformation 
structure and derive a corresponding parallelization scheme.

\subsection{Algebraic Characterization of Polar Transformations}
For a permutation $\pi \in \Sym(n)$ acting on the coordinate indices of a BLBC, 
we let $\pi$ act on a codeword by $\pi(\c)=\c\P_\pi$, 
and on a matrix by $\pi(\A)=\A\P_\pi$, 
where $\P_\pi\in\mbF_2^{n\times n}$ is the 
matrix representation of a permutation  $\pi$.

 \begin{definition} (\textit{Polar Transformation Parameter.})
 
  For a BLBC $\mcC_b$, the polar transformation of~\cite{lin2020unidec}
  can be parameterized by a permutation 
  $\pi \in \mathrm{Sym}(n)$ 
  that maps $\mcC_b$ to a polar subcode $\mcC_{\polar}$,
  \begin{equation}
  \pi: \mcC_\BLBC \rightarrow \mcC_\polar.
  \end{equation}
  We refer to $\pi$ as the \textbf{transformation parameter} of the polar transformation.
 \end{definition}
    The proof of Proposition~\ref{prop:PD} can therefore be reformulated as follows.   
  Consider the dynamic frozen constraint as  
\begin{equation}
\W_\pi = \pi(\G_\BLBC)\G_\polar^{-1},
\end{equation}
and the set $\mcC_\polar := \{\c_\polar = \pi(\c_\BLBC): \c_\BLBC \in \mcC_\BLBC\}$ 
forms a polar subcode $\{\c_\polar = \u\G_\polar: \u \in \mcR(\W_\pi)\}$. 
To see this, take an $\m \in \mbF_2^k$ and let $\u = \m\W_\pi \in \mcR(\W_\pi)$, then
\begin{equation}
\c_p = \u\G_\polar = \m \pi(\G_\BLBC)\G_\polar^{-1}\G_\polar = \m\G_b\P_\pi = \pi(\c_\BLBC).
\end{equation}
This gives an algebraic characterization of Proposition~\ref{prop:PD}.

To determine when different transformation parameters share the same decoder, 
we characterize their induced dynamic frozen constraints. 
 \begin{definition}(\textit{Dynamic Frozen Constraint Induced by a Polar Transformation}.)
  The dynamic frozen constraint induced by the polar transformation of 
  $\mcC_b$ with parameter $\pi$ is given by the mapping
  \begin{equation}
  \phi: \mathrm{Sym}(n) \rightarrow \mbF_2^{k \times n},\quad
  \phi(\pi) = \M_\pi,
  \end{equation}
  where 
  \begin{equation}
  \M_\pi= \mathrm{RREF}(\pi(\G_b)\G_p^{-1}).
  \end{equation}
  The row-reduced form 
  $\M_\pi$
  preserves the row space of $\W_\pi$, 
  and hence the induced linear constraint on $\u$, 
  while yielding a SC-compatible constraint representation.
  
 \end{definition}

  The mapping $\phi$ captures the dependence of the 
dynamic frozen constraint on the transformation parameter.
However, no known closed-form relation exists
between the transformation parameter $\pi$ 
and the resulting information set $\phi(\pi)$, 
making direct optimization over $|\Sym(n)| = n!$ intractable. 

The authors of~\cite{lin2020unidec} circumvented this by deriving 
explicit constructions for eBCH codes; 
for general BLBCs,~\cite{lin2025unidec} resorted to AI-assisted search. 
Both lines of work focus on finding a single optimal parameter. 
Extending these approaches to parallel decoding would require 
finding multiple parameters independently.
Other than the further enlarged search space, 
each transformation parameters also potentially requires different dynamic frozen constraintsi, 
and hence different decoder designs, 
which further complicates the parallelization. 

In contrast, 
rather than searching for multiple compatible parameters, 
we study the algebraic structure of $\phi$,
in particular its \textbf{equivalence classes},
to identify families of parameters that share a common decoder by construction.

In the following we denote
\(
  \mathrm{Aut}(\mcC)= \{a \in \mathrm{Sym}(n): a(\mcC) = \mcC\}
\)
as the automorphism group of a code $\mcC$, and the identity element of the group 
as $\text{id}$ which for any $\c \in \mcC$, $\text{id}(\c) = \c$.
We next prove that the action of $\mathrm{Aut}(\mcC_b)$ on the transformation parameter 
$\pi\in \mathrm{Sym}(n)$ induces 
equivalent polar transformations.

\begin{theorem}\textit{(Invariance of $\phi$.)}
\label{thrm:invariantDFclass}
Let $\mcC_b$ be a BLBC, and let $\phi(\pi)$ denote the dynamic frozen constraint 
induced by the transformation parameter $\pi \in \Sym(n)$.
For any $a \in \mathrm{Aut}(\mcC_b)$,
\[
\phi(  \pi\circ a ) = \phi(\pi).
\]
\end{theorem}

\begin{IEEEproof}
Since $a \in \mathrm{Aut}(C_b)$, there exists $\E \in GL_k(\mathbb{F}_2)$ such that
\(
 \G_b \P_a = \E \G_b.
\)
Hence,
\begin{equation}
\G_b \P_{(\pi \circ a )} 
= \G_b \P_{a} \P_{\pi} 
= (\E \G_b) \P_{\pi} 
= \E (\G_b \P_{\pi}).
\end{equation}
Therefore,
\begin{equation}
\mcR(\G_b \P_{(\pi \circ a )} \G_p^{-1})
=
\mcR(\G_b \P_{\pi} \G_p^{-1}).
\end{equation}
Since RREF is the canonical representative of rowspaces,
\begin{equation}
\phi(\pi \circ a ) = \phi(\pi).
\end{equation}
Therefore, all parameters in the right coset $\pi \text{Aut}(\mcC_b)$ 
induce the same dynamic frozen constraint.
\end{IEEEproof}

Notice that the coset $ \pi\mathrm{Aut}(\mathcal C_b)$ 
is generally neither $ \mathrm{Aut}(\mathcal C_b)$
nor the automorphism group of the resulting polar subcode. 
If we directly use an automorphism $a \in \mathrm{Aut}(\mathcal C_b)$ as the 
transformation parameter $\pi$, it effectively collapses Lin's method back to Trifonov's 
fixed frozen set design since $\phi(a) = \phi(\text{id})$.

\begin{corollary} \textit{(Equivalent Classes of Polar Transformations.)}
  Let $\pi\text{Aut}(\mcC_b):= \{\pi \circ a: a \in \text{Aut}(\mcC_b)\}$ be the right coset with representative $\pi$.
Using any element $\sigma \in \pi \text{Aut}(\mcC_b)$ 
as the transformation parameter, 
\begin{equation}
  \sigma : \mcC_b \rightarrow \mcC_p,
\end{equation}  
since $\phi(\pi) = \phi(\sigma)$ as we proved in Theorem~\ref{thrm:invariantDFclass}, 
they yields the same image polar subcode
\begin{equation}
 \mcC_p = \{\c_p = \u\G_p: \u \in \rowspan(\phi(\pi))\}.
\end{equation}  
Identical dynamic frozen constraints therefore 
induce equivalent polar transformations.
\end{corollary}

\begin{remark}[{\it Theoretical Interpretation}]
To enable a direct parallelization of the framework proposed in~\cite{lin2020unidec},
Theorem~\ref{thrm:invariantDFclass}
shifts the action from the codeword space to the transformation parameter $\pi$.
The well-characterized automorphism groups of classical BLBCs directly furnish equivalent polar 
transformations without code-specific redesign or per-permutation decoder reconfiguration.
Unlike prior AED methods, the automorphism action is on the \textbf{parameter side} instead of the codeword side.

\end{remark}

\vspace{-0.6em}
\begin{algorithm}[h]
\caption{Polar Ensemble Decoding for BLBCs}
\label{alg:POD}
\KwInput{
Channel output $\mathbf{y}$, base permutation $\pi$, 
parallelization path size $M \le |\text{Aut}(\mcC_b)|$,
automorphism subset $\{a_1,\ldots,a_M\} \subseteq \text{Aut}(\mcC_b)$,
RREF Elimination matrices $\{\E_1, \ldots, \E_M\}$
}

\For{$i=1$ \KwTo $M$}{
  $\mathbf{y}_i \leftarrow (\pi\circ a_i(\mathbf{y}))$\;
  $\hat{\mathbf{m}}_{p,i} \leftarrow \mathrm{PolarDec}(\mathbf{y}_i)$\;
}

$j \leftarrow \mathrm{Combiner}\big(\{\hat{\mathbf{m}}_{p,i}\}_{i=1}^M\big)$\;

$\hat{\mathbf{m}} \leftarrow \hat{\mathbf{m}}_{p,j}\mathbf{E}_j$\;
\KwOutput{Estimated message $\hat{\mathbf{m}}$}
\end{algorithm}

\begin{figure*}[!t]
\centering
\resizebox{\textwidth}{!}{%
\begin{tikzpicture}[
  >=Latex,
  font=\footnotesize,
  node distance=6mm and 7mm,
  line/.style={-Latex, thick},
  ghost/.style={draw, thick, dotted, minimum height=7mm, minimum width=7mm, inner sep=1.0pt, align=center},
  block/.style={draw, thick, minimum height=7mm, minimum width=13mm, inner sep=1.0pt, align=center},
  bigblock/.style={draw, thick, minimum height=8mm, minimum width=16mm, inner sep=1.0pt, align=center},
]

\coordinate (m0) at (0,0);

\node[ghost, right=of m0] (Einv) {$\E^{-1}$};
\node[ghost, right=of Einv] (E) {$\E$};
\node[bigblock, right=of E] (blbcenc) {BLBC enc};
\node[ghost, right=of blbcenc] (Pinv1) {$\pi$};
\node[ghost, right=of Pinv1] (Gninv) {$\G_p^{-1}$};
\node[ghost, right=of Gninv] (Gn) {$\G_p$};
\node[ghost, right=of Gn] (P) {$\pi^{-1}$};


\node[block, right= of P] (channel){Channel};
\coordinate (chsplit) at ($(channel.east)+(3mm,0)$);


\node[block, right= of chsplit, yshift=+10mm] (Prx) {$\pi\circ a_1$};
\node[block, right= of chsplit] (Prxh1) {$\pi\circ a_2$};
\node[right= of chsplit , yshift=-10mm] (vdots) {$\vdots$};
\node[block, right= of chsplit , yshift=-20mm] (Prxhm) {$\pi\circ a_M$};

\node[bigblock, right=of Prx] (polardec) {Polar dec};

\node[bigblock, right=of Prxh1] (polardec1) {Polar dec};

\node[bigblock, right=of Prxhm] (polardecm) {Polar dec};

\node[block, right=of polardec1] (chcomb) {Combiner};
\node[block, right=of chcomb] (Eout) {$\hat{\m}_{p,j}\E_j$};
\draw[line] (m0) -- node[above, yshift=1mm] {$\m$} (Einv);
\draw[line] (Einv) -- node[above, yshift=1mm] {$\m_p$} (E);
\draw[line] (E) -- (blbcenc);
\draw[line] (blbcenc) -- node[above, yshift=1mm] {$\c_b$} (Pinv1);

\draw[line] (Pinv1) -- node[above, yshift=1mm] {$\u$} (Gninv);
\draw[line] (Gninv) -- (Gn);
\draw[line] (Gn) -- node[above, yshift=1mm] {$\c_p$} (P);
\draw[line] (P) -- node[above, yshift=1mm] {$\c_b$} (channel);

\draw[thin] (channel.east) -- (chsplit);
\draw[line] (chsplit)|-(Prx);
\draw[line] (chsplit)--(Prxh1);
\draw[line] (chsplit)|-(Prxhm);
\draw[line] (Prx) -- (polardec);
\draw[line] (Prxh1) -- (polardec1);
\draw[line] (Prxhm) -- (polardecm);
\draw[line] (polardec) -| node[above, yshift=1mm] {$\hat \m_{p,1}$} (chcomb.north);
\draw[line] (polardec1) -- node[above, yshift=1mm] {$\hat \m_{p,2}$} (chcomb.west);
\draw[line] (polardecm) -| node[below, yshift=-1mm] {$\hat \m_{p,M}$} (chcomb.south);
\draw[line] (chcomb) --  node[above, yshift=1mm]{$j$}(Eout) ;
\draw[line] (Eout) --  node[right, xshift=3mm]{$\hat\m$}($(Eout.east) + (5mm, 0)$) ;


\end{tikzpicture}%
}
\caption{Overall $\pod$ architecture. All the dashed blocks are virtual components. 
}
\label{fig:df_arch}
\end{figure*}
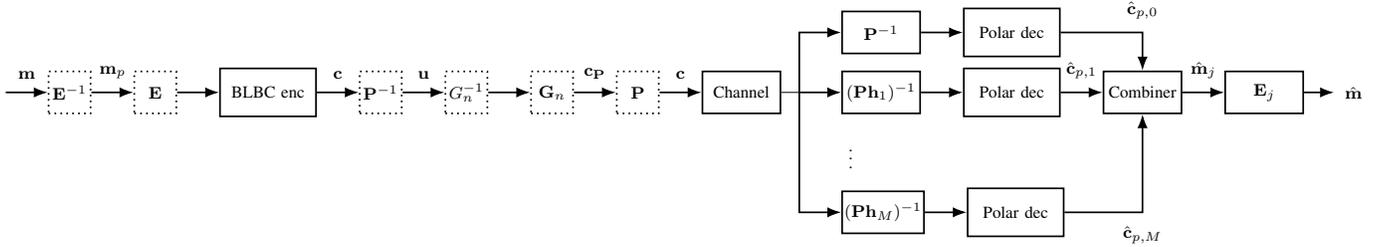

\vspace{-1.18em}
\subsection{Proposed Algorithm}
The decoding procedure of $\pod$ is summarized in Algorithm~\ref{alg:POD} 
{and depicted in Fig.~\ref{fig:df_arch}}. 
To exploit the invariant class formed by the right-coset, we sample $M$ different 
elements from the automorphism group,
and apply them to the base permutation parameter.
We then run independent decoding paths over the resulting ensemble of polar transformations.
The polar decoders across all $M$ paths admit the same structure thanks to the invariant dynamic frozen design $\phi(\pi \circ a) = \phi(\pi)$.
To recover the original BLBC message, the selected polar-domain estimate
$\hat{\m}_{j}$ is mapped back using the corresponding elimination matrix
$\E_j$, i.e., $\hat{\m}=\hat{\m}_{j}\E_j$.

Finally, we combine the candidates from all decoding paths. Since each candidate admits both polar-subcode and BLBC interpretations, the combiner may either (i) select the candidate with the best polar-decoder path metric, or (ii) apply the original BLBC parity-check and select a valid decoded candidate. The combiner then outputs the index $j$ of the selected path.

Our parallel design enables the reduction of SCL decoding latency while 
retaining the same effective list size 
by varying the number of permutations $M$,
introducing design flexibility in latency and hardware complexity.

\subsection{LTA-type SC-Invariant Transform Analysis}

We now analyze the SC-invariancy of polar transformation and 
justify our current uniform sampling strategy.

\begin{definition}[{\it Effective Difference Between Two Transformations}]
Suppose we fix a polar transformation of $\mcC_b$ 
with base parameter $\pi$. 
The effective difference between the base transformation $\pi$ and the coset transformation $(\pi \circ a) \in \pi \text{Aut}(\mcC_b)$ 
is 
  \(
  \pi\circ a \circ \pi^{-1}.
  \)
To see this, note that for any $\c_b \in \mcC_b$,
\begin{equation}
 \pi\circ a \circ \pi^{-1}(\pi(\c_b)) = \c_b\P_\pi\P_\pi^{-1}\P_a\P_\pi = \pi\circ a(\c_b). 
\end{equation}
The set of all such effective differences forms the conjugate of the automorphism group by $\pi$,
\begin{equation}
  \pi\text{Aut}(\mcC_b)\pi^{-1}.
\end{equation}
\end{definition}

Although there has been discussion on larger classes of SC-invariant permutations~\cite{ye2022SC_polar_complete}, 
we focus on LTA here to have a quick and clean bound.
If 
\(
  \pi\circ a \circ \pi^{-1} \in \text{LTA},
\)
then the transformation is SC-invariant relative to the base permutation. 
Hence, the set of LTA-type SC-invariant transformations is
  \(
  \pi\text{Aut}(\mcC_b) \pi^{-1} \cap \text{LTA}.
  \)

In the following we provide an upper bound on the probability that 
a randomly sampled coset element results in an LTA-type SC-invariant transformation, 
and consequently a lower bound on the probability that all $M$ samples are SC-variant.
\begin{theorem}[{\it Probability of sampling an LTA-type SC-invariant transformation}]
  \label{thrm:2m_SC_prob}
  Consider an $(n=2^m, k)$ BLBC $\mcC_b$. 
  Let $|\text{Aut}(\mcC_b)| = s_o 2^{s_e} $ where $s_o$ is odd.
  The intersection of the conjugate subgroup $\pi \text{Aut}(\mcC_b)\pi^{-1}$ 
  and LTA$(m)$ 
  satisfies
  \begin{equation}
    |\pi\text{Aut}(\mcC_b)\pi^{-1} \cap \text{LTA}(m)| \leq 2^{s_e}.
  \end{equation}
    
  The probability $p_\text{LTA}$ that a uniformly sampled coset element lies in $\text{LTA}(m)$ is therefore upper bounded by
  \begin{equation}
    p_\text{LTA} 
    = \frac{|\pi\text{Aut}(\mcC_b)\pi^{-1} \cap \text{LTA}(m)|}{|\text{Aut}(\mcC_b)|}
    \leq \frac{2^{s_e}}{|\text{Aut}(\mcC_b)|} = \frac{1}{s_o}.
  \end{equation}
  Note that this bound holds for all $\pi \in \text{Sym}(n)$ and for general BLBCs with arbitrary $\text{Aut}(\mcC_b)$.
  Although loose, this upper bound is quite general.
\end{theorem}

\begin{IEEEproof}
  LTA$(m)$ is a 2-group since it is of order $2^{m(m+1)/2}$. 
  Hence, the intersection is a 2-subgroup of the conjugate subgroup,
  and is therefore contained in a Sylow 2-subgroup of $\pi|\text{Aut}(\mcC_b)|\pi^{-1}$.
  Since $|\text{Aut}(\mcC_b)| = s_o 2^{s_e}$,  
  a Sylow 2-subgroup is of order $2^{s_e}$~\cite{artin1991algebra}. 

  The multi-kernel case analysis, 
  including a generalization of the UPO and LTA frameworks to code lengths 
  that are not powers of two, 
  is left for future work.
\end{IEEEproof}

In the worst case, the entire Sylow 2-subgroup lies in $\text{LTA}(m)$. 
Thus, the event that all $M$ samples are pairwise SC-variant
is equivalent to the event that they lie in distinct cosets of 
${\pi\text{Aut}(\mcC_b)\pi^{-1} \cap \text{LTA}(m)}$ in $ {\pi\text{Aut}(\mcC_b)\pi^{-1}}$.
The probability of this event, denoted by $p_M$, is therefore lower bounded by
\begin{equation}
  p_M 
  \geq \prod_{i=0}^{M-1}(1-\frac{i}{s_o})
 \simeq 1 - \frac{M(M-1)}{2s_o}.
\end{equation} 

\begin{example}[$p_\text{LTA}$ of eBCH codes]
 For $(n=2^m, k)$ eBCH codes, the automorphism is the semilinear affine group 
 \begin{equation}
  |\text{A}\Gamma\text{L}(1, 2^m)| = 2^m(2^m -1)m.
 \end{equation}
 We split $m = m_o 2^{m_e}$ where $m_o$ is an odd number.
Then $s_o = (2^m-1)m_o$ and
\begin{equation}
  p_\text{LTA} \leq \frac{1}{(2^m-1)m_o},
\end{equation}
which decays exponentially with $m$. 
\end{example}

\section{Simulation Results}\label{sec:simulation}
\begin{figure}[!t]
	\centering 
		\includegraphics[width= .9\columnwidth]{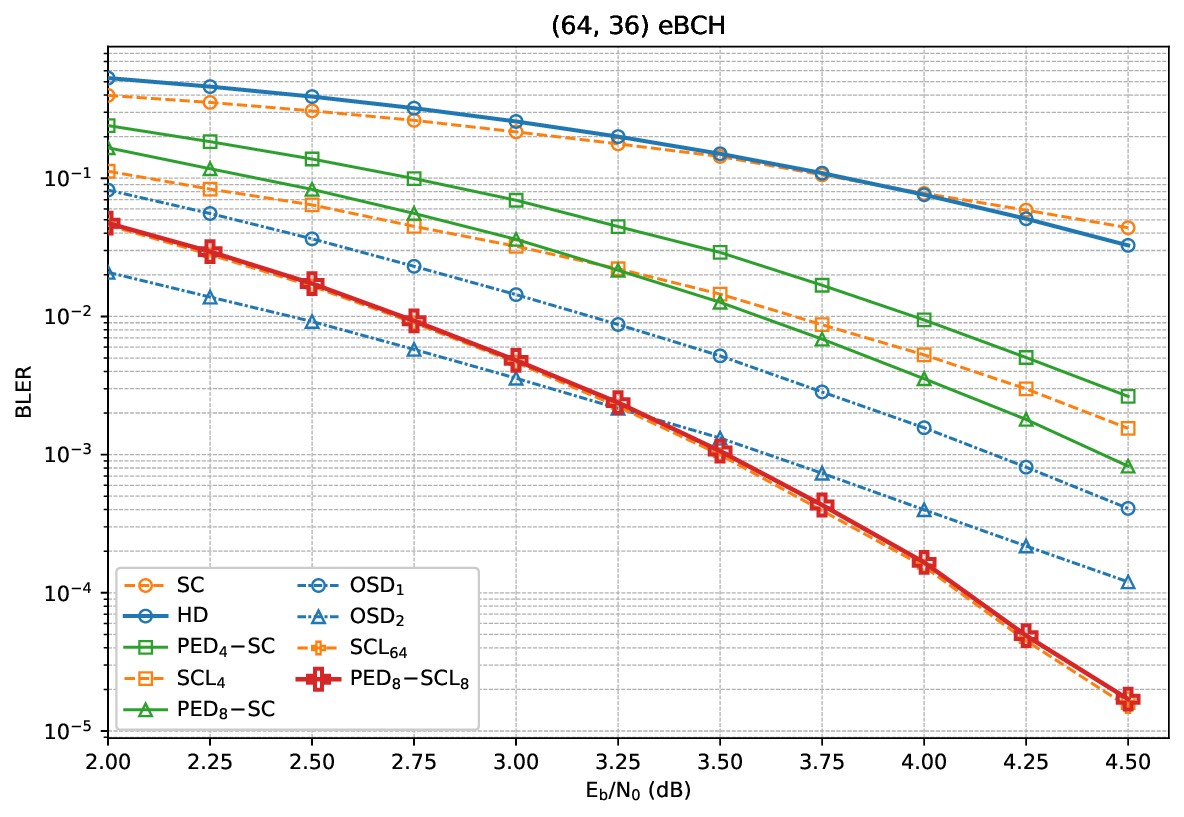}
    \caption{Performance of eBCH (64, 36) code.}
\label{fig:eBCH_6_5}
\end{figure}

\begin{figure}[!t]
	\centering 
		\includegraphics[width= .9\columnwidth]{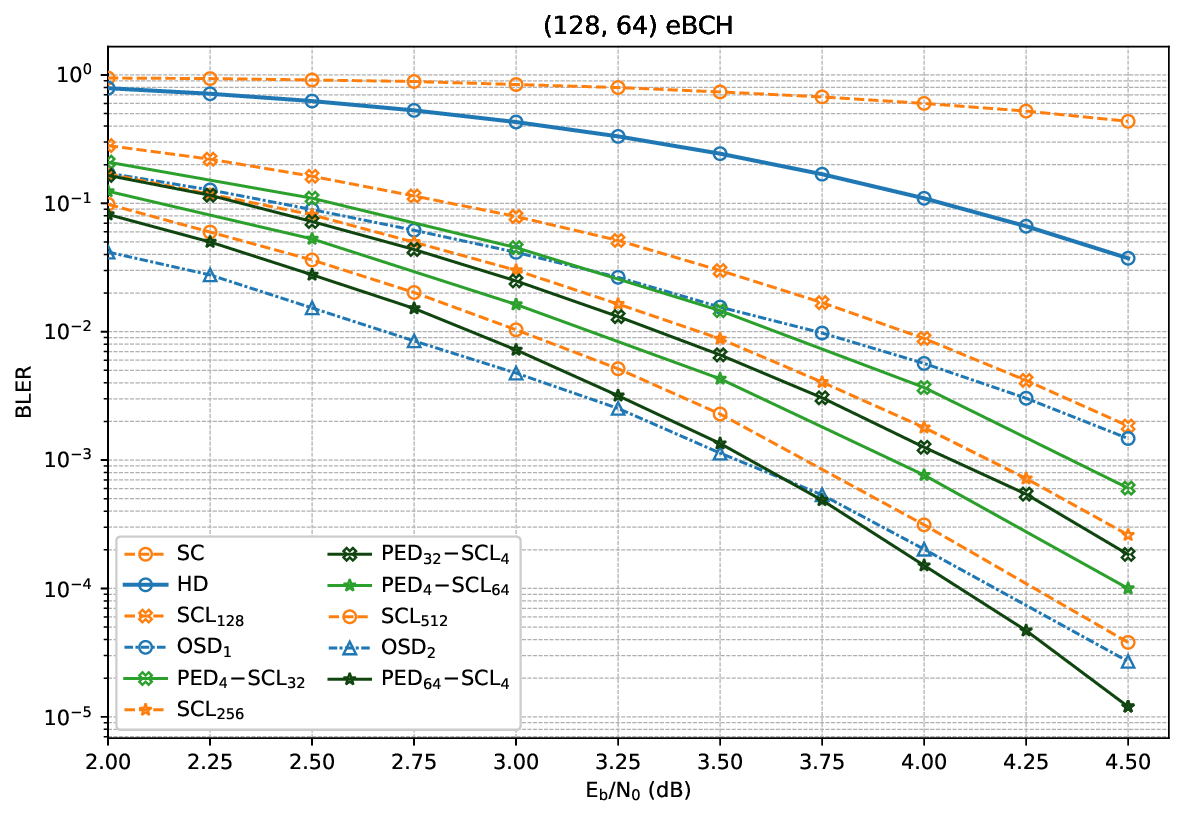}
    \caption{Performance of (128, 64) eBCH code.
    }
\label{fig:eBCH_7_10}
\end{figure}

{
In this section, we present simulation results over the AWGN channel to demonstrate the design flexibility of $\pod$. 
For comparison, we include soft-decision ML, SC, and SCL decoding, as well as hard-decision (HD) decoding based on the theoretical error probability of the corresponding algebraic decoding algorithm for each BLBC~\cite{lin2001ECC}. 
We compare the block error rate (BLER) performance of these decoding methods. 
In the legend, $\pod_M$ denotes $\pod$ with $M$ permutations drawn from the automorphism coset associated with the base parameter $\pi$ found in~\cite{lin2020unidec}, followed by $\mathsf{SC}$ or $\mathsf{SCL}_L$, where $L$ denotes the list size of the polar decoder.

We use the base parameter obtained by the methods in~\cite{lin2020unidec} to transform each BLBC into a polar subcode. 
We generate the automorphism group using the Schreier-Sims algorithm and store it in Base and Strong Generating Set (BSGS) form~\cite{seress2003permutation}. 
This allows us to uniformly sample $M$ distinct elements from $\Aut(\mcC_b)$ in polynomial time.
The two combiner designs described in Section~\ref{sec:POD} exhibit similar performance experimentally. 
In the following, we present the results obtained using path-metric comparison as the combiner.

We first consider the $(64,36)$ extended BCH code in Fig.~\ref{fig:eBCH_6_5}. 
When $\sc$ is used as the polar decoding module within $\pod$, the BLER improves monotonically with the number of permutations and rapidly approaches ML performance. 
In particular, when $M=8$ and $L=8$, the resulting performance is comparable to that of $\mathsf{SCL}_{64}$, while benefiting from reduced decoding latency through parallelization.

Fig.~\ref{fig:eBCH_7_10}
demonstrates the scalability of $\pod$ to longer codes. 
For the same effective list size, 
$\pod_{64}$-$\scl_4$ outperforms $\pod_4$-$\scl_{64}$ by approximately $0.5$ dB, 
indicating the diversity gain provided by $\pod$. 
Moreover, against $\scl_{512}$, 
the parallel $\pod_{64}$-$\scl_4$ achieves better BLER 
while using only half the effective list size. 
These results suggest that the diversity brought by $\pod$ 
can be more effective than simply increasing the $\scl$ candidate list size,
since $\pod$ exploits structural diversity that pure list expansion cannot capture.

\begin{figure}[!t]
	\centering 
		\includegraphics[width= .9\columnwidth]{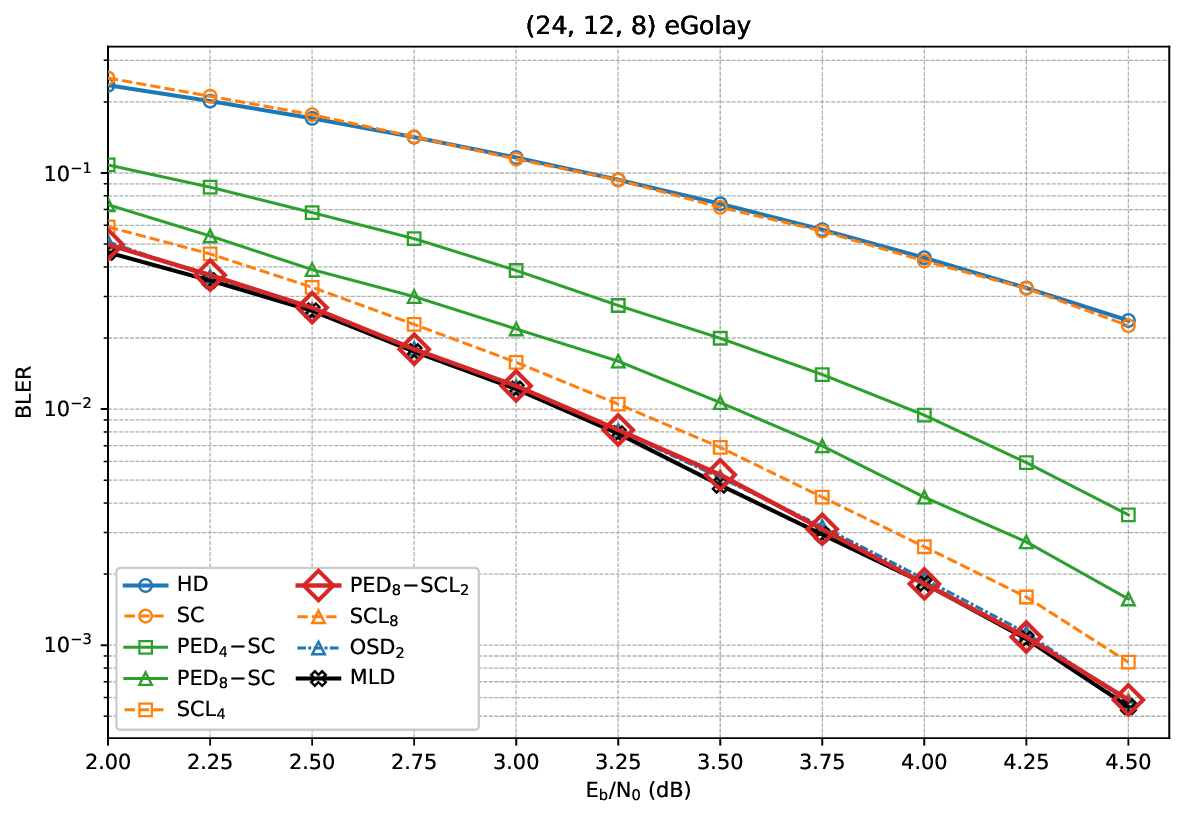}
    \caption{Performance of (24, 12) eGolay code.
    }
    \vspace{-5pt}
\label{fig:eGolay_24_12}
\end{figure}
We further illustrate the generality of $\pod$ using the $(24,12)$ extended Golay code. 
Its automorphism group is the Mathieu group $M_{24}$~\cite{BERLEKAMP197140}. 
Since $n=24$ is not a power of two, we adopt the multi-kernel polar code~\cite{garby2017multiker_polar} as $\G_p$ in this experiment:
\begin{equation}
\mathbf{G}_p = 
\begin{bmatrix} 1 & 0 & 0 \\ 1 & 0 & 1 \\ 1 & 1 & 1 \end{bmatrix}
\otimes
\begin{bmatrix} 1 & 0 \\ 1 & 1 \end{bmatrix}
\otimes
\begin{bmatrix} 1 & 0 \\ 1 & 1 \end{bmatrix}
\otimes
\begin{bmatrix} 1 & 0 \\ 1 & 1 \end{bmatrix}.
\end{equation}

Fig.~\ref{fig:eGolay_24_12} shows that $\pod$ improves substantially over plain $\mathsf{SC}$ decoding and achieves nearly identical performance to SCL decoding at the same effective list size. 
With $M=8$ and $L=2$, $\pod$ approaches ML performance and is comparable to both $\mathsf{SCL}_{64}$ and $\mathsf{OSD}_2$, while offering lower decoding latency. 
The experiments demonstrate the performance gain provided by $\pod$.
This suggests that the random coset sampling remains effective beyond the power-of-two setting.

Overall, these results show that 
$\pod$ achieves flexible latency-parallelism tradeoffs
across BLBC families with nontrivial automorphisms. 
By operating on transformation parameters rather than codewords, 
PED enables a unified polar transformation framework 
that exploits automorphisms for parallelization 
across different code families.

}

\section{Conclusions}\label{sec:conclude}
In this work, we proposed a parallelization of the   
parameterized universal polar decoding framework for arbitrary BLBCs 
with nontrivial automorphism. 
By exploiting the automorphism structure of the embedded BLBC, 
$\pod$ enabled efficient parallel decoding within the polar framework. 
Decoding within the automorphism right coset induced equivalent polar subcodes 
that were decoded independently using standard polar-style decoders. 
$\pod$ thus provides scalable tradeoffs among decoding performance, latency, 
and implementation complexity, establishing a versatile solution for parallel soft-decoding of BLBCs.

Future work will extend $\pod$ to enhanced polar transformations~\cite{lin2025unidec}, 
analyze the selection of $M$ transform parameters in the right coset,
and generalize the framework to non-binary linear block codes.

\clearpage\balance
\bibliographystyle{IEEEtran}
\bibliography{ref}

\end{document}